\documentclass[twocolumn, 10pt, nofootinbib, final]{revtex4} \graphicspath{{./pic/}}
\usepackage{array}
\usepackage{numprint}
\usepackage{xspace}

\npdecimalsign{.}

{\begin{figure}[t]\begin{center}\noindent}{\end{center}\end{figure}} \newenvironment{widepict}%
{\begin{figure*}\begin{center}\noindent}{\end{center}\end{figure*}} \def\look#1{(see Fig.~\ref{#1})}

\makeatletter \def\rmpt@#1.#2{#1#2} \def\put@pt#1#2.#3{\ifx#3\empty#2#1\else
    #2.\kern-.25em\relax#1\rmpt@#3\fi} % magnitude
\def\m#1{\ensuremath{\put@pt{^m}#1.\empty}} % degrees
\def\arcdeg#1{\ensuremath{\put@pt{^\circ}#1.\empty}} % arc minutes
\def\arcmin#1{\ensuremath{\put@pt{'}#1.\empty}} % arc seconds
\def\arcsec#1{\ensuremath{\put@pt{''}#1.\empty}} 
\newcommand{\frc}[2]{\raisebox{2pt}{$#1$}\big/\raisebox{-3pt}{$#2$}} \makeatother
\def\np#1{\numprint{#1}}
\def\e{\ensuremath{\mathrm{e}^{-}}\xspace}
\def\QE{\ensuremath{\langle\mathrm{QE}\rangle}\xspace}

\begin{document} \title{ESPriF: the Echelle Spectropolarimeter of the BTA Primary Focus. Correction of
        Low-Frequency Variations in the Star Image}

\author{M.V.~Yushkin, E.V.~Emelianov, Yu.B.~Verich, and V.E.~Panchuk} \affiliation{Special Astrophysical
    Observatory} \begin{abstract} The development of a corrector for low-frequency variations in the star image at
    the input of ESPriF,  the echelle spectropolarimeter of the BTA primary focus, is reported. New
    technical solutions have made it possible to extend the operating frequency range to 10\,Hz for stars brighter
    than \m{13}. \end{abstract} \maketitle

\section{Introduction}
During development of a new spectral device for the BTA, ESPriF~\cite{proj2020}, the
main requirement was to increase the efficiency, both in comparison with previous designs and to compensate
for the process of performance degradation of the BTA optics, operated since the late 1970s. In addition to
reducing Fresnel losses on ESPriF lens optics special attention was paid to compensation for low-frequency
variations in wavefront tilts. Let us list the developments that preceded this.

The first works are devoted to compensation of wavefront tilts in order to increase the limiting magnitude of
spectral equipment. Since~2000, a local wavefront tilt corrector, LTC~\cite{loccorr2001}, has been
operating at the Nasmyth focus of the 6-m BTA telescope, which facilitates the work of the observing
astronomer and provides profit in throughput by one magnitude on the spectrographs NES~\cite{hires2017} and
MSS~\cite{mss2014}. In~2020, the development of a new local corrector of the BTA Nasmyth focus, named the
``Wavefront Tilt Compensator'', has begun. At the same time the rich statistics of the LTC operation are
taken into account, which allowed, in particular, to detect the anisotropy of image fluctuations caused by
wind load on the telescope structure~\cite{lowfr2022}.

During the reconstruction of the echelle spectrograph of the BTA primary focus PFES~\cite{probres1997}
in~2005, a local primary focus corrector, LPFC~\cite{progapp2006}, was tested, which showed less inertia
compared to the LTC (due to the reduced mass of the plane-parallel plate and the mass of the corresponding
mechanics). Currently, the mechanics of the LPFC has been completely redesigned taking into account changes in
the optical design and dimensions of the primary focus echelle spectropolarimeter, as well as a new approach
to the problem of compensating for wavefront tilts.

In 2017, a local corrector for the fiber-optic input of the echelle spectrograph of a meter-class telescope
(LCMT) was developed~\cite{design2015}. This made it possible to stabilize the illumination of the optical
fiber without forming a conjugate pupil at the entrance end face. The results of the development of
high-resolution spectral equipment with a fiber-optic input for telescopes of moderate diameter made it
possible to integrate a small-sized wavefront curvature corrector into the system for matching the apertures
of the telescope and optical fiber.

The listed systems operate at relative apertures $A = 1 : 30$, $1 : 4$ and $1 : 13$, respectively. Moreover,
the concept of the LCMT, developed for the Cassegrain focus ($A = 1 : 13$) of a 1-meter telescope, served as
the basis for the development of a new corrector for the primary focus of the BTA ($A = 1 : 4$) for the
high-resolution echelle spectropolarimeter ESPriF.

The next stage in creating wavefront correctors includes not only increasing the limiting magnitude of
spectral equipment, but also increasing the spectral resolution with a simultaneous increase in throughput,
which contradicts the ``canonical'' principles of high-resolution spectroscopy, when the product of throughput
(L) and spectral resolution (R) is invariant. This approach requires not only more accurate compensation of
wavefront tilts, but also an improvement in the image quality itself, i.e. compensation of higher orders of
wave aberrations. As a first step in this direction, in~2020, the LCMT scheme was supplemented with a
wavefront curvature corrector, which allows real-time compensation for defocusing of the telescope's optical
system caused by both atmospheric turbulence in the surface layer and movements of air masses in the dome
space. Currently, a similar system for compensating for wavefront curvature has been developed for the primary
focus of the BTA ($A = 1 : 4$), where it is being tested. This paper also examines the prospects for creating
a system for correcting wavefront curvature for foci with a different relative aperture.

\begin{widepict}
    \includegraphics[width=\textwidth]{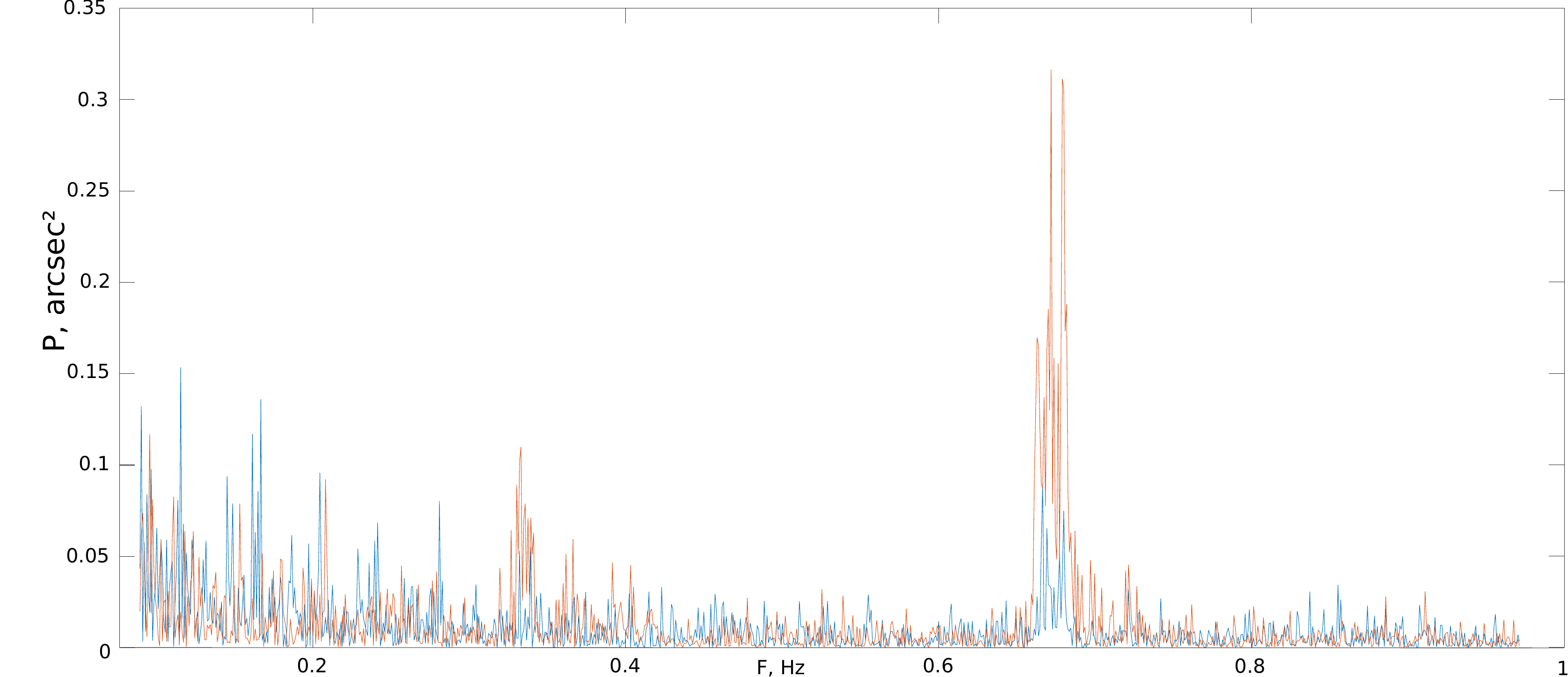}
    \caption{Spectra of image oscillations along the A (red) and Z (blue) axes.}
    \label{AZhighfreq}
\end{widepict}

\section{Characteristics of low-frequency image vibrations}
The work Klochkova et al.~\cite{improv2020} notes the prevailing frequencies of image oscillations identified
over the years of operation of the BTA: in the range of $0.7\div0.9\,$Hz on both axes, in the range of
$0.8\div1\,$Hz for the azimuth axis and $0.4\,$Hz for the zenith distance axis. When position of an object is
corrected by influences on the telescope drives, oscillations arise with a decay time of about 25\,s. In
May~2023, after the installation of new frequency converters of the BTA axis motors, again, studies were
carried out on the accuracy of object tracking. For this purpose, several rows of images of bright stars were
obtained with short exposures ($50\div200\,$ms), the accuracy of calculating the barycentre of the star image
in a single frame was~\arcsec{0.05}.

Frequency analysis showed the presence of several vibration harmonics~\look{AZhighfreq}. The maximum
occurs in oscillations with a frequency of about $0.66\,$Hz (period of about $1.5\,$s) for both axes,
although along the azimuthal axis (A) oscillations with such a period are more pronounced than along the
zenith distance axis (Z). The second peak occurs at a frequency of $0.33\,$Hz (period $3\,$s) and is
observed only in the power spectrum of oscillations along the A~axis. In the region of oscillations with
periods of more than 5\,s, a ``picket fence'' of peaks is observed in the power spectrum, apparently
associated with the absence of obvious harmonic oscillations, although the amplitude of such deviations
can be significant. Along the A axis, the relatively long-period ($10\div50\,$s) deviation of the star
image from the average position reached \arcsec{4}, which is most likely caused by wind load. The
deviation of the star image from the average position along the Z axis is more stochastic.

In addition to objective statistics, the employees of the astrospectroscopy
laboratory~\footnote{\url{https://www.sao.ru/hq/ssl/}} have also accumulated subjective experience over
decades of observations at the BTA. In particular, it is known that the characteristics of fluctuations in
the position of images are not necessarily related to fluctuations in their shape: sometimes the diameter
of a stably positioned image changes noticeably, up to the separation of the whole into fragments. In
other words, even with good images, low-frequency correction of wavefront tilts and curvature is
necessary.

\begin{widepict}
    \includegraphics[width=\textwidth]{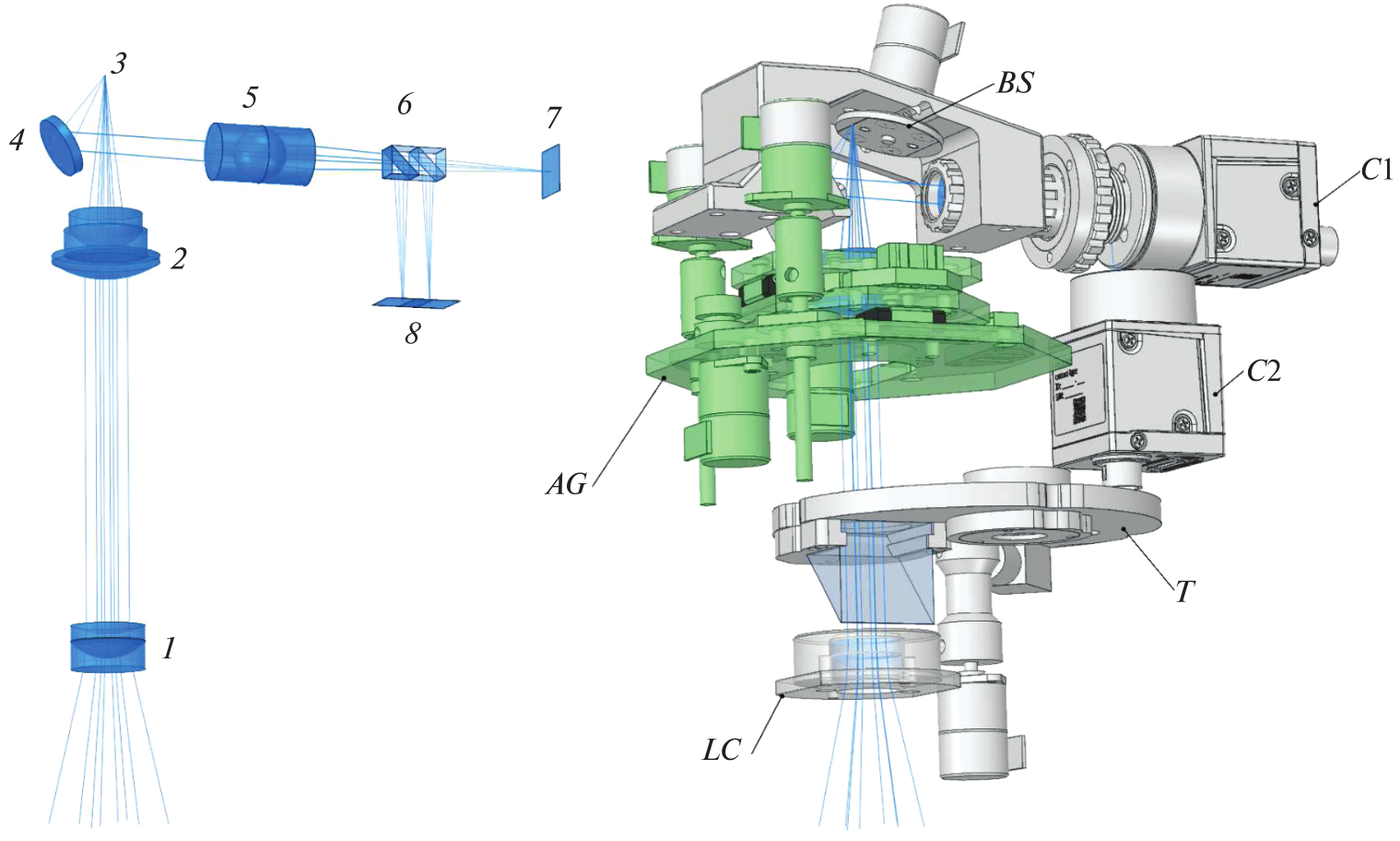}
    \caption{Optical design of the afocal reducer. The focusing element (lens, left) can be moved along three orthogonal axes to
        compensate for wavefront curvature and tilts.}
    \label{esprifin}
\end{widepict}

\section{The scheme of the  of low-frequency image oscillations corrector for ESPriF}
The ESPriF echelle spectropolarimeter was developed over several years, with forced breaks. For example,
during attempts carried out in 2018--2019 to complete the process of replacing the working main mirror (MM) of
the BTA (MM No.\,2) with an updated mirror (MM No.\,1), it turned out (due to the removal of a significant
layer of glass from surface of MM No.\,1) that the maximum extension of the surface of the primary focus (PF)
beyond the mounting flange of the turntable in the PF cabin has significantly decreased. Thereby, in some
modes of the SCORPIO device~\cite{sco2005} already used in routine observations the position of the BTA
focusing device was close to the limit switch, and some modes (polarimetric) turned out to be inaccessible at
all, which during further operation of the MM No.\,1 on the telescope would require reconstruction of the
so-called BTA PF adapter, which performs the functions of calibration and guiding. In our project, we limited
ourselves to manufacturing a temporary assembly of the pre-slit unit of ESPriF, with a set of functions that
allows us to evaluate the limiting magnitude of the spectrograph. At the same time, the function of viewing a
large field had to be transferred to a small ($D = 20\,$cm) catadioptric telescope, which was being
reconstructed to replace the general guide of the BTA.

After the MM No.\,2 was returned to the telescope, part of the volume intended to accommodate the nodes of the
pre-slit unit and reduced due to a change in the flange focal distance was freed up, and a new design for the
entrance to the spectrograph was developed.

An optical calculation was performed of the new pre-slit unit of ESPriF, which includes the following
elements~\look{esprifin}:
\begin{enumerate}
    \item afocal reducer~--- lens system (elements ``1'' and~``2'') converting a converging beam of rays from
the main mirror of the BTA into a converging beam with the same aperture, but with the ability of quick
defocusing compensation by shifting one optical element~``2''; the same moving element also provides
compensation for the wavefront tilt moving along two orthogonal axes in a plane perpendicular to the
optical axis of the telescope;
    \item the entrance slit viewer with a defocus detector~--- projection lens with beam splitting
cubes in a converging beam (elements ``5'' and~``6'', respectively), camera ``C1'' fixing the position of
the barycenter of the star image and camera ``C2'', which records postfocal and prefocal monochromatic
images to control focus position;
    \item polarimetric module, including two super-achromatic phase-shifting plates~--- quarter-wave and
half-wave, located on the turret~``T'', as well as a plaque Savard (not shown in the figure) with the
possibility of inserting the latter behind the entrance slit of the spectrograph with simultaneous
installation of a defocus compensator;
    \item a calibration channel for illuminating the entrance slit of the spectrograph with line and
continuous spectrum sources; Fig.~\ref{esprifin} shows a total internal reflection prism on the
turret~``T'' for inputting calibration light into the beam.
\end{enumerate}

The idea of an afocal reducer goes back to the design of a lens collimator moved along orthogonal axes to
compensate for fluctuations in the image of a star. This scheme was first proposed in the design of a slitless
stellar spectrograph with guiding and interferometric spectrum references, included in the first generation of
the BTA equipment~\cite{30years2006}.

The functions of the new pre-slit unit include: control of pointing to the object, selection of the decker
used, image capture by the guiding system, continuous and line spectrum calibrations, swapping polarization
analyser plates, changing plate rotation angles, compensating for wavefront tilts and curvature. The nodes
performing all these functions were combined into the so-called input unit (IU), including nodes of the
pre-slit unit and a node of slits (deckers).

Let us show that the development of a corrector consisting of low-inertia units and parts transfers the problem
of advancing to the region of relatively high frequencies from the class of optomechanical to the class of
optoelectronic problems. To achieve this, some energy estimates were performed.

\section{Energy calculations for the BTA primary focus}
The ESPriF efficiency estimation based on real observations showed that at the phase
of pilot operation of the spectrograph, point sources with a magnitude of up to~\m{15} in the V~band with a
resolution of $R = \np{15000}$ will be available for observation. Consequently, for satisfactory functioning of
the image corrector, the photodetector of the viewing channel must register such objects with an exposure time
not exceeding $0.1\,$s. Let's perform an energy calculation for the viewing channel, taking into account the
design features of the ESPriF input unit.

The effective area of the BTA main mirror, $S_{eff}$, taking into account the obscuration by the primary
focus cabin and the reflectivity of main mirror surface is $S_{eff} = 0.8\pi (600^2-180^2 )/4 \approx\np{2e5}\,\text{cm}^2$. To
estimate the number of photons collected by the telescope at the primary focus,
we take Vega ($\alpha\,$Lyr) as the zero magnitude standard. Let us perform the calculation separately for
each of the four photometric bands U-B-V-R${}_\mathrm{c}$. The spectral flux density of Vega's radiation
behind the atmosphere, $f_\lambda^{Vega}$, the effective wavelength, $\lambda_{eff}$, and the width,
$\Delta\lambda$, of each photometric band are taken from the work of Fukugita et al.~\cite{Fukugita1995}
(Table~1). We will estimate the photon flux $F_{ph}^{15^m}$ at the telescope focus from a star of~\m{15}
integrally in each filter based on the following relation:

\begin{equation*}
F_{ph}^{15^m} = \tau_\lambda\frac{f_\lambda^{Vega}\Delta\lambda}{e_\lambda^{ph}}S_{eff}\cdot10^{-6},
%\label{Fph15}
\end{equation*}
where $e_\lambda^{ph}$ is the energy of a single photon, $\tau_\lambda$ is the
atmospheric absorption coefficient in a given photometric band, and the multiplier $10^{-6}$ is the ratio of the
\m{15} and \m{0} fluxes. The coefficients $\tau\lambda$ are calculated using the formula:

\begin{equation*}
\tau\lambda = 10^{-0.4K_\lambda X},
\end{equation*}
where $K_\lambda$ is the atmospheric extinction coefficient, and $X = \sec Z$ is the air mass. We will assume
that the light from the star passes through an air mass of $X = 2$ (zenith distance $Z = \arcdeg{60}$),
observations are rarely carried out at larger zenith distances. Atmospheric extinction coefficients for the
BTA location ($K_U = \m{0.615}$, $K_B = \m{0.319}$, $K_V = \m{0.184}$, $K_R = \m{0.114}$) taken
from~\cite{Neizv1983}. Table~\ref{tab1} shows the results of calculating the photon flux $F_{ph}^{15^m}$ in
the primary focus of the BTA from the star of \m{15} in each photometric band. It can be seen that the total
flux in the entire optical range (bottom line of Table~\ref{tab1}) will be about \np{4.7e5}~photons per
second.

The viewing channel receives part of the star image that did not pass inside the spectrograph and is reflected
from the slit blades. ESPriF entrance slits are laser cut into a polished stainless steel plate. We will
consider the reflection coefficient from a polished steel plate equal to~$0.6$. To obtain a spectral
resolution of \np{15000}, \np{22500} and \np{30000}, slits with a width of $120\,\mu$m ($\arcsec{1}$), $90\,\mu$m
(\arcsec{0.75}) and $60\,\mu$m (\arcsec{0.5}) respectively, can be installed at the input to the spectrograph.
The height of all three slits is $240\,\mu$m, which is equivalent to \arcsec{2}. Let's evaluate the efficiency
of the viewing channel with the widest slit ($R = \np{15000}$). In the case of the spectrograph operating in a
mode with higher spectral resolution, the amount of light in the viewing channel will be greater, both due to
a decrease in the fraction of light passing inside the spectrograph, and due to an increase in the brightness
of the observed object, since with increasing spectral resolution the limiting magnitude of ESPriF will
decrease.

%\newcolumntype{P}[1]{>{\centering\arraybackslash}m{#1}}
\def\hb#1#2{\hbox to #1{\hss #2 \hss}}
\begin{table*}\def\arraystretch{1.5}
\begin{tabular}{|c|n{5}{0}|n{5}{1}|n{3}{2}|n{2}{2}|n{2}{2}|n{7}{1}|n{6}{0}|n{2}{3}|n{6}{1}|n{6}{1}|}
\hline
Band & {\parbox{1cm}{$\lambda_{eff}^{Vega}$, \AA}} & {{$\Delta\lambda$, \AA}} & \parbox{2cm}{$f_\lambda^{Vega}$,
$\strut\frc{\mathrm{erg}}{\mathrm{cm^2 s \text{\AA}}}$} &
{\hb{1.8cm}{$e_\lambda^{ph}$}} & {\hb{8mm}{$\tau_\lambda$}} & {\parbox{1.3cm}{$F^{15^m}_{ph}$, ph\,s${}^{-1}$}} &
{\parbox{1.3cm}{$F_{ph}^{Moon}$, ph\,s${}^{-1}$}} &
{\hb{9mm}{\QE}} & {\parbox{1.3cm}{$S^*$, \e{}s${}^{-1}$}} &
{\parbox{1.3cm}{$S^{Moon}$, \e s${}^{-1}$}}\\
\hline
U & 3709 & 526 & 4.28e-9 & 5.36e-12 & 0.32 & 26881 & 9258 & 0.25 & 1142 & 6221 \\
B & 4393 & 1008 & 6.19e-9 & 4.52e-12 & 0.56 & 154608 & 28224 & 0.64 & 16821 & 48554\\
V & 5439 & 827 & 3.60e-9 & 3.65e-12 & 0.71 & 115825 & 16540 & 0.66 & 12996 & 29343 \\
R${}_\mathrm{c}$ & 6410 & 1568 & 2.15e-9 & 3.10e-12 & 0.81 & 176172 & 25402 & 0.22 & 6589 & 15022\\
\hline
\textbf{Full} & \multicolumn{5}{c|}{} & 473486 & 79424 & & 37548 & 99140 \\
\hline
\end{tabular}
\caption{Estimation of radiation flux in the ESPriF viewing channel in different photometric bands.}
\label{tab1}
\end{table*}

Many years of experience in monitoring of the astroclimatic conditions at the BTA location shows that the
average diameter of the turbulent disk of the star at half maximum, so called seeing parameter, is
\arcsec{1.5}~\cite{astrocl2011}. Determining the position of the star image at the ESPriF input slit is performed
by calculating the coordinates of the center of gravity in a part of the frame twice the diameter of the
turbulent disk. Therefore, with average image quality, we will read an area of size
$\arcsec{3}\times\arcsec{3}$. Diego~\cite{Diego1985} calculated the energy concentration at different seeing
conditions. Using these calculations we find that under average seeing conditions for the BTA location 40\%
of the stellar radiation flux gathered by the telescope will pass through the ESPriF slit of size
$\arcsec{1}\times\arcsec{2}$, and the area $\arcsec{3}\times\arcsec{3}$ will pass 84\% of total energy. Thus,
44\% of the total energy of the star arriving at the primary focus of the BTA will be sent to the viewing
channel to determine the coordinates of the centroid. Taking into account the reflection coefficient from the
slit blades equal to~$0.6$ (position~``3'' in Fig.~\ref{esprifin}), the reflection coefficient of the folding
mirror as~$0.8$ (position~``4'') and the transmittance of the viewing channel lens and the beam splitter equal
to~$0.8$ (positions~``5'' and~``6'' respectively), the overall efficiency of the viewing channel is equal
approximately to~$0.17$. The viewing channel camera (C1), used to monitor the position of the star image, is
equipped with a SONY IMX249 CMOS sensor (position~``7''). The average quantum efficiency of IMX249,
\QE, in the corresponding photometric band is given in a separate column of Table~\ref{tab1}.
Taking into account this quantum efficiency, the signal from the star, $S^*$, in each band, read from the
sensor in area $\arcsec{3}\times\arcsec{3}$, was calculated. It can be seen that the integral signal (bottom
line of Table~\ref{tab1}) is \np{3.7e4}~photoelectrons per second. Therefore, in a single frame with an
exposure time of 0.1\,s, there will be approximately \np{3700} photoelectrons per image of a star of \m{15}.
Without considering the intrinsic noise of the radiation receiver, the signal-to-noise ratio on a single frame
in the approximation of Poisson statistics, defined as $S/N = \sqrt{S^*}$, is approximately equal to~60, which
in the first approximation is sufficient to determine the coordinates of the image centroid with accuracy
\arcsec{0.1}. However, in real conditions we must consider both the background and the intrinsic noise of the
detector.

The source of background illumination is the Earth's atmosphere, which has both its own glow and scatters
moonlight (we do not take into account anthropogenic factor such as artificial skyglow, aerosol concentration
and so on, although their contribution increases every year). The implementation of spectroscopic
observational programs at the BTA on stellar topics, as a rule, occurs on bright nights, so the main
contribution to the background illumination will be made by scattered light from the Moon. Detailed studies of
the sky background at the BTA location during lunar phases close to the full Moon were not carried out, since
all photometric programs are carried out in the dark. Therefore, to estimate the level of scattered light, we
will use the atmospheric model developed for the European Southern Observatory~\cite{Jones2013}. Skyglow
calculations at different wavelengths for various ESO observatories are available through the
SKYCALC~\footnote{\url{https://www.eso.org/observing/etc/skycalc}} web service. For the online calculator we
used the following input parameters: observatory~--- ``La Silla'' (\np{2400}\,m above sea level), air mass~---
$X = \sec Z = 2$, Moon phase~--- $1.0$~(full), height of the Moon above the horizon~---~$\arcdeg{45}$,
distance from the object to the Moon~---~$\arcdeg{45}$. Table~\ref{tab1} shows separately in each photometric
band the calculated fluxes of a scattered moonlight radiation, $F_{ph}^{Moon}$, at the telescope focus coming
from the sky area in one~square arcsecond. On average, on a moonlit night \np{7.9e4}\,photons of scattered
light from one square arcsecond of sky will arrive at the focus of the telescope every second. It is clear
that this value will vary greatly both from the height of the Moon above the horizon, the distance of the
object from the Moon, and from the state of the atmosphere as a whole (the amount of water vapour, the amount
of aerosols, the presence of clouds, etc.). The overall efficiency of the viewing channel when observing
extended sources consists of the reflection and transmittance coefficients of optical elements (``3'', ``4'',
``5'' and ``6'' in Fig.~\ref{esprifin}) and is $0.38$. As noted above, to determine the coordinates of the
centroid with average image quality we read the area $\arcsec{3}\times\arcsec{3}$, thus the area from which
scattered light will be collected (minus the area of the input spectrograph slit) is 7\,arcsec${}^2$. Taking
into account the average quantum efficiency of the viewing channel detector~\QE in each photometric band we
can estimate the integral signal from the background illumination, $S^{Moon}$. As a result, we find that the
detector area occupied by the star's image accounts for a total of \np{9.9e4}~photoelectrons per second, which
is almost three times the signal from the star. In this case, $S/N$~can be estimated using the formula:

\begin{equation*}
S/N=\frac{S^*}{\sqrt{S^*+S^{Moon}}}.
\end{equation*}
In our case, the signal-to-noise ratio on a single frame with an exposure of $0.1$\,s in the viewing channel
when observing a \m{15}~star will be $S/N \approx 30$, which is noticeably worse, but still sufficient to
confidently recognize an object against the sky.

In reality, the star image occupies a large number of sensor pixels, and $S/N$~calculated for the integral
flux only indicates the accuracy of the intensity measurement. Since we are interested in the accuracy of
positional measurements, we must use a different criterion. Namely, we will assume that to determine the
position of the centroid with an accuracy of~\arcsec{0.1} we need to have a spatial resolution element of no
more than~\arcsec{0.5} and $S/N\ge5$ per element of spatial resolution. The pixel size of the SONY IMX249
detector is $5.86\,\mu$m, which in projection onto the celestial sphere corresponds to a scale of
\arcsec{0.05}\,pix${}^{-1}$. Thus, there are 400~pixels per area of the sky in one square arcsecond, and the
image of a star with the central part cut out will occupy an area of \np{2800}~pixels. It follows from this
that there will be slightly more than one photoelectron per unit detector element of the viewing channel
during an exposure time of 0.1\,s.

The signal-to-noise ratio, taking into account the detector intrinsic noise, can be calculated using the
following formula:
\begin{equation*}
S/N=\frac{S^*}{\sqrt{S^* + S^{Moon} + n_r^2P + DP}},
\end{equation*}
where $n_r$ is the readout noise, $P$~is the number of pixels occupied by the star image, $D$~is the dark
current. The dark current of the IMX249 detector is $0.03\,$\e per pixel per second, and can be
neglected. The readout noise varies depending on the operating mode of the detector and averages 7\,\e.
It turns out that on a single frame we will have $S/N < 0.2$ per pixel. The easiest way to increase the
signal-to-noise ratio is binning, or combining several pixels. With CMOS detectors, unlike CCD, binning of the
pixels is possible only in a firmware of the camera (off-chip binning), which increases~$S/N$ by only a factor
equal to the square root of the number of binned pixels. To satisfy the above spatial resolution criterion, we
can average (or sum) the signal over an area of $10\times10\,$pixels, thereby increasing $S/N$ by a factor
of~10. However, as a result, for a star of \m{15} with an exposure of 0.1\,s we obtain $S/N < 2$ per element
of spatial resolution, which is not enough for the required accuracy of the image corrector.

Based on the presented calculations, we find that with a CMOS detector, compensation of the natural
frequencies of the BTA oscillation is possible only when observing stars with ESPriF up to~\m{13}.
Consequently, compensation for wavefront tilts caused by atmospheric phenomena and turbulent air movements in
the space under the dome is possible only for even brighter objects.

Let us now estimate the maximum possible S/N for a star of~\m{15}, which we could obtain by changing the image
scale by 10~times (matching the size of the spatial resolution element with the size of the receiver element)
or using another type of detector that allows combine pixels directly on the chip (on-chip binning) as
happens with CCDs so that each element of spatial resolution is read with an error of $n_r = 7\,\e$.
According to Table~\ref{tab1} to an area of size $\arcsec{0.5}\times\arcsec{0.5}$ on the light receiver of the
viewing channel during an exposure time of 0.1\,s $S^*=134\,\e$ comes on average from the star,
$S^{Moon}=354\,\e$ from the background illumination. As a result, we find that when matching the image scale
with the detector resolution element, the maximum possible signal-to-noise ratio for a star of~\m{15} in the
viewing channel is $S/N \approx 6$.

We come to the conclusion that compensation for wavefront tilts caused by telescope oscillations at natural
frequencies is only possible when observing stars up to~\m{15} in the case of changing the type of
photodetector or reducing the image scale in the viewing channel. When observing fainter objects, the image
corrector will operate in auto-guiding mode. Changing the optical design of the pre-slit unit of ESPriF by
installing a beam splitter in front of the entrance slit of the spectrograph will not lead to significant
increase in the limiting magnitude of the viewing channel. The beam splitter allows us to send 10\%~of the
radiation flux from the full image of the star to the viewing channel, so we can determine the coordinates of
the centroid not from the remnants of the image with the central part cut out, but from the brighter central
part. As the above calculations showed, the efficiency of the viewing channel in the current design of the
pre-slit unit of ESPriF is~$0.17$ for a poit sources under average weather conditions and~$0.38$ for extended
sources, and in the case of using a beam splitter, the efficiency of the viewing channel will not
exceed~$0.10$, but it will be the same as for point and extended sources. This circumstance will increase the
contrast of the image and, in the case of ideal weather conditions, advance to~\m{16} provided that the
spatial resolution element is matched with the pixel (or ``superpixel'') size. But given that ESPriF plans to
observe even fainter objects (up to~\m{19}) after upgrading the spectrograph camera, it is necessary to
develop a drastically new guiding system, possibly based on offset stars.

In our project, the frequency of correction of the slopes and curvature of the wavefront is determined by:
\begin{enumerate}
\item the magnitude of the observed object,
\item inertia of the moving element in the afocal reducer,
\item the duration of image analysis and performing the appropriate movements.
\end{enumerate}

Leaving the low correction frequency as a free parameter in spectroscopy of very faint objects we plan to use
the combined guiding method, implemented on spectrographs of the Nasmyth-2 focus~\cite{loccorr2001}. In this
method mismatches that change monotonically are eliminated by influencing the drives of the telescope control
system and quasiperiodic mismatches are eliminated by acting on the corrector drives.

\section{Algorithmic support of the corrector}
The developed low-frequency wavefront corrector is an integral part of the ESPriF spectrograph, therefore the
corrector control system is one of the spectrograph control subsystems.

The basis of the control system is a mini-computer running Gentoo GNU\slash Linux OS, located directly in the hanging
part of the spectrograph. The unified instrument control interface is a graphical shell for several system
daemons: obtaining scientific images, obtaining images from the slit\slash deckers viewing cam- era,
controlling the spectrograph nodes, and control- ling the low-frequency wavefront corrector.

All control daemons for individual spectrograph nodes exchange information through either local TCP\slash IP
sockets or UNIX sockets. The interaction of the remote user with them is provided by the network daemon which
opens network TCP\slash IP sockets protected by TLS as well as protected web sockets for the web interface.

Receiving images from the viewing camera is an independent operation. The last received image is contained in
a buffer, which can be transmitted to any client upon a network request: for visual display or for calculating
the coordinates of the star centroid. In automatic correction mode a separate application, corrector
daemon, controls the gain level and exposure of the CMOS detector and also sets the area of interest (to
increase the speed of reading and data transfer). Typically, the area of interest coincides with the limiting
zone in which auto-correction by mechanical movements of the lens unit is available. After calculating the
centroid coordinates the required amount of lens displacement is determined to compensate for the
corresponding wavefront tilt. This value is calculated by a software PID controller.

The control module for stepper motors and other devices of the spectrograph is also designed as a separate
network daemon. Taking the necessary displacement values, the demon tries to complete the task, and if the
limit of movement of the corrector lens is reached, it reports the impossibility of further movement. In this
case, the corrector daemon moves the telescope drives (in correction mode) so that the current position of the
image is in the center of the permissible area of lens movement. Damped oscillations established as a result
of correction of the object's position by the telescope drives are compensated by displacements of the lens
unit (element ``2'' in Fig.~\ref{esprifin}).

The image processing algorithm is as follows. After reading the next image a histogram is constructed from
which (at the inflection point after the first maximum) the background level is determined. Based on the level
of the background calculated in this way a binary mask image is constructed (one corresponds to a signal
above the background level). Next, morphological operations are performed to clean up the noise,
four-connected regions are found and numbered. Each area is checked for the aspect ratio of the describing
rectangle and the occupied area (they must lie within the specified limits). If after this more than one
object is still detected the objects are sorted by brightness or distance to the target coordinates (the
choice of algorithm lies with the user). Next, the found displacements of the centroid coordinates pass
through the software PID controller and the displacement of the correction lens is determined, which is
transmitted via a network socket to the control daemon of the mechanical components of the device or to the
telescope drives (if the local correction limit is reached).

After the first installation of the device on the telescope, as well as after any manipulations that could
cause a change in the position of the corrector assembly, it is calibrated. Calibration determines the
position of the displacement axes of the corrective lens relative to the image axes, as well as the
orientation of the displacement vectors carried out by the telescope drives. And if the first part,
calibration of the local corrector, can be done in laboratory conditions using an artificial star, then
calibration of the correction axes by a telescope is performed only using a real celestial object. In the
general case, the axes of movement of the lens can be non-orthogonal and the displacement value may represent
a nonlinear dependence on the angle of rotation of the stepper motor rotor (for example, when controlling the
position of the lens with cam mechanisms). The following procedures are performed:

\begin{enumerate}
\item the telescope is pointed at a given object;
\item both coordinates are set to the middle position;
\item by correcting the telescope the user places the star as close as possible to the center of the
correction area and accurately focuses the star image;
\item with a step of~$1/10$ of the entire range of movement, displacements are performed first along the
$U$~axis and then along the $V$~axis of the corrector, then the coefficients of the polynomial for converting
the position of the shaft of the corrector motors into image displacement are determined, as well as the
direction vectors of the corrector axes in the image;
\item similarly, the image is shifted by the telescope drives (but only two frames are performed for each
coordinate) to determine the orientation of its axes on the frame, then this position is determined taking
into account the current positional angle.
\end{enumerate}

Thus, the setup and operation of the low-frequency corrector of the ESPriF spectrograph turned out to be more
complex than working with the local Nasmyth focus corrector (LNC).

\section{Corrector hardware}
The pre-slit part of the spectrograph contains a significant number of moving units: both an auto- corrector
unit and various mechanisms for changing slits (deckers), filters, polarizing elements, etc. Therefore, we
will briefly describe the hardware of the control system. To work with stepper motors that ensure the movement
of the listed components, a control module has been developed based on the STM32F303VBT6 microcontroller,
the module can either be connected to the control computer via USB or work as part of a CAN bus. To focus the
spectrograph camera lens, based on the STM32F103C6T6 microcontroller, a control unit has been developed (by
reverse engineering) that connects to a computer via USB or as part of a CAN bus. Line and continuous spectra
(thorium-argon and halogen lamps, respectively) are formed by a separate device developed on the basis of an
STM32F072C8T6 microcontroller, also connected to a computer via USB or as part of a CAN bus. The
glow mode of a hollow cathode lamp (Th+Ar) is controlled by an adjustable constant current source. High voltage
(up to 200\,V) is generated by a boost DC-DC converter. Since the hanged spectrograph will need frequent
calibration with a line spectrum, attention is paid to minimizing the exposure time of the calibration
spectrum.

\section*{Conclusions}
Experience in the development and operation of low-frequency star image correctors at relative aper- tures
of~$1 : 30$ (Nasmyth focus of the BTA) and~$1 : 13$ (the 1-m~telescope) made it possible to build a corrector
for the tilt and curvature of the wavefront for the primary focus of the BTA~($1 : 4$). Tests of the corrector
at the telescope confirmed the correctness of the basic technical solutions. The inertia of mechanical drives,
which determines the limit of correction frequencies in the Nasmyth-2 focus, has been structurally overcome in
the new device. Numerical estimates have shown that for most objects whose spectroscopy requires image
correction, the upper frequency limit is determined not by the mechanical properties of the corrector, but by
the sensitivity of the object registration system and the features of the algorithm for generating executive
commands. We plan to use the found technical solutions on other spectral devices.

\section*{Acknowledgments}
Observations with the 6-meter telescope of the Special Astrophysical Observatory of the Russian Academy of
Sciences were supported by the Ministry of Science and Higher Education of the Russian Federation.

\section*{Funding}
Financial support was provided under the grant No.\,075-15-2022-262 ``Multi-wave study of non- stationary
processes in the Universe'' (13.MNPMU.21.0003).

\section*{Conflict of interest}
The authors declare no conflict of interest.

\newcommand{\ab}{Astrophysical Bulletin }
\newcommand{\alet}{Astronomy Letters }
\newcommand{\pasp}{Publ. Astron. Soc. Pacific }

\end{document}